\documentclass[preprint2,twoside,dvips]{aastex}
\usepackage{graphicx}
\newcommand{\kms}{\mbox{ km s}^{-1}}        
       
\begin{document}        
\title{SUPERNOVA REMNANTS IN THE SEDOV EXPANSION PHASE: THERMAL X-RAY EMISSION}
\author{Kazimierz J.~Borkowski, William J. Lyerly, and Stephen P.~Reynolds}
\affil{Department of Physics, North Carolina State         
University, Raleigh, NC 27695}        

\begin{abstract}        
Improved calculations of X-ray spectra for supernova remnants (SNRs) in the
Sedov-Taylor phase are reported, which for the first time include reliable
atomic data for Fe L-shell lines. This new set of Sedov models also allows
for a partial collisionless heating of electrons at the blast wave and for
energy transfer from ions to electrons through Coulomb collisions. X-ray
emission calculations are based on the updated Hamilton-Sarazin spectral
model. The
calculated X-ray spectra are succesfully interpreted in terms of three
distribution functions: the electron temperature and ionization timescale
distributions, and the ionization timescale averaged electron temperature
distribution. The comparison of Sedov models with a frequently used single 
nonequilibrium ionization (NEI)
timescale model reveals that this simple model is generally not an appropriate
approximation to X-ray spectra of SNRs. We find instead that plane-parallel
shocks provide a useful approximation to X-ray spectra of SNRs, particularly 
for young SNRs. Sedov X-ray models described here, together with simpler 
plane shock and single ionization timescale models, have been implemented
as standard models in the widely used XSPEC v11 spectral software package.
\end{abstract}        
        
\section{INTRODUCTION}

X-ray astronomy has advanced significantly in recent years because of
improved X-ray instrumentation. Several recent X-ray satellites
provided complementary capabilities: high spatial resolution and
mapping of the entire sky (ROSAT), spatially-resolved spectroscopy
(ASCA), and broad-band coverage ({\it Beppo}SAX). In the field of
supernova remnants (SNRs), this has led to many advances, such as the
discovery of the nonthermal synchrotron X-ray emission produced by
energetic electrons accelerated at a shock front, identification of a
number of new neutron stars and associated synchrotron nebulae,
discovery of new SNRs, and identification of young, ejecta-dominated
SNRs.  Spatially-resolved X-ray spectroscopy, provided by the ASCA
satellite, played a crucial role in these discoveries. The ASCA
database now contains a large number of excellent spectra of
SNRs. Observations with the next generation of X-ray satellites ({\it
Chandra} and XMM-{\it Newton}) have begun to provide even better
quality data, ensuring continuing progress in the SNR research.

While the new X-ray observations led to significant progress, X-ray
spectral data on SNRs, and ASCA data in particular, have been mostly
underutilized. The fundamental reason for this unsatisfactory
situation is that X-ray emitting plasmas in SNRs are not in ionization
equilibrium, significantly complicating analysis of their X-ray
spectra. Until recently most X-ray observers have not had access even
to simple nonequilibrium ionization (NEI) models. The simplest NEI
model consists of an impulsively heated, uniform and homogeneous gas,
initially cold and neutral. While this constant temperature,
single-ionization timescale model is often used in analysis of X-ray
spectra of SNRs, in most cases it is unlikely to be a good
approximation to shock-heated plasmas in SNRs. Because X-ray spectra
are sensitive to the detailed structure of a SNR, the best description
of SNR spectra is through hydrodynamical modeling, coupled with X-ray
emission calculations. This approach is obviously impractical in most
cases. That leaves idealized hydrodynamical structures, such as
self-similar hydrodynamical solutions, as the best choice for general
NEI models. The Sedov-Taylor self-similar solution, describing a point
blast explosion in a uniform ambient medium, is the most well-known
and useful of such structures.

X-ray emission calculations for the Sedov models have been performed numerous
times in the past, with the most extensive set of calculations presented by
Hamilton, Sarazin, \&\ Chevalier (1983). The shape of the X-ray spectrum, 
spatially integrated over the whole remnant, depends on the shock speed $v_s$, 
the ionization timescale $\tau_0$ (which we define as the product of the 
remnant's age $t_0$ and the postshock electron number density $n_{es}$), 
and
the extent of electron heating at the shock front. In addition, as typical for 
most X-ray emitting thermal plasmas, line strengths and to a lesser extent 
continua depend on heavy-element abundances. Sedov models have been 
recalculated using updated atomic data (Kaastra \&\ Jansen 1993) and 
successfully applied to modeling of X-ray spectra of SNRs in the Large 
Magellanic Cloud (Hughes, Hayashi, \&\ Koyama 1998).

The most uncertain aspect of X-ray modeling in the framework of the 
Sedov model is the extent of electron heating in a blast wave. Because of the
low densities typical for the interstellar medium (ISM), shocks in the 
ISM are collisionless, 
where the ordered ion kinetic energy is dissipated into random
thermal motions through
collisionless interactions involving magnetic fields. Electrons, whose
collisions with ions produce the observed X-ray emission from SNRs, are most 
likely also heated to some 
degree in collisionless shocks, but the details of this process are not well 
understood.  In the extreme case of a very high Mach-number shock,
in the shock frame both electrons and ions enter with the shock
velocity $v_s$, so that after randomization one expects $kT_i \sim 
m_i v_s^2$ and $k T_e \sim m_e v_s^2$, that is, $T_e/T_i \sim 1/1836$,
unless some collisionless heating of electrons (at the expense of ions)
also occurs.
Hamilton et al. (1983) considered two extreme possibilities,
either full electron--ion equilibration at the shock front, or electron
heating only through Coulomb collisions in the postshock gas, without any
collisionless heating at the front itself. Hughes et al. (1998) used these two 
limiting classes
of models in their modeling of LMC SNRs, and found better (but not perfect) 
agreement for models without any collisionless heating. While these results 
suggest that the full electron--ion equipartition might not hold in SNRs, the
extent of collisionless electron heating remains an open question. 

These questions are critical for observations since both the X-ray
continuum and lines originate from electrons: from electron
bremsstrahlung or electron impact excitation of ions.  Until X-ray
observations are able to resolve thermal Doppler widths of emission
lines, ion energy remains invisible except as a pool from which to
heat electrons.  The detailed shape of the electron distribution
affects both continuum and line emission.  This introduces another 
potential problem:
the possible deviation of the electron energy distribution function
from a Maxwellian distribution, in particular the presence of a
high-energy tail. Electrons in such a tail might modify the
high-energy X-ray spectrum of a SNR through interactions with ions
and magnetic fields.

Theoretical modeling of collisionless shocks, performed with the purpose of 
accounting for the observed intensity and properties of Galactic cosmic rays,
revealed that energetic particles generated within the shock should strongly
modify the shock structure. Such cosmic-ray modified shocks transfer 
a significant amount of kinetic energy into cosmic rays, which then diffuse
upstream of the original shock and partially decelerate incoming gas in 
a region ahead of it (the shock precursor). The temperature of the thermal
gas is much lower than in a standard (nonmodified) shock with the same
velocity, because of this efficient transfer of kinetic energy into cosmic
rays. This will affect X-rays produced by thermal plasmas in cosmic-ray 
modified shocks, because of strong dependence of X-ray emission on gas
temperature. The modeling of modified shocks has been usually done in the
framework of steady-state, plane-parallel shocks, preventing a direct
comparison with the Sedov model discussed above. Such a comparison is now
possible, as time-dependent, cosmic-ray modified shock models have been
constructed recently by Berezhko, Yelshin, \&\ Ksenofontov (1994, 1996).

The continuing improvement in theory just described is not followed by 
a similar improvement in the X-ray data analysis.
Instead of using Sedov models (and even more sophisticated
hydrodynamical models, including those with cosmic-ray modified
shocks), many observers still rely on the ionization equilibrium models,
which are generally not acceptable for SNRs. Sometimes spectra are
modeled by Gaussian lines and free-free continua, a purely 
phenomenological approach which does not allow for determination of physical
parameters in X-ray emitting plasma. The frequently used constant 
temperature, single ionization timescale  NEI model is better than
an equilibrium plasma model, but it is still not appropriate for SNRs.
In order to make full advantage of current and future X-ray data
on SNRs. a concentrated effort is required to close the gap between the 
relatively sophisticated theory and the current crude data analysis.
The theoretical work described here focuses on a most basic set of models
which is necessary for analysis of X-ray data, and which should be routinely
used by X-ray observers.
We present new calculations of spatially-integrated X-ray emission from Sedov 
models in \S~2, allowing for partial electron heating at a shock front. 
A preliminary version of this work was presented in Lyerly et al.~(1997).
In
\S~3, we discuss our results in terms of suitably chosen temperature
and ionization timescale distribution functions, and 
compare Sedov models with simpler NEI models such as a single timescale,
constant temperature model or a plane-parallel shock. The sensitivity of
X-ray spectra to model parameters is discussed in \S~4. 
We present a summary of our results in \S~5.

\section{X-RAY EMISSION CALCULATIONS} 

\subsection{Sedov-Taylor Dynamics}
We base our spectral calculations on a Sedov-Taylor (ST)
blast wave model.  The ST model begins with a spherically symmetric point 
blast expanding adiabatically into a uniform ambient medium with the 
adiabatic index $\gamma=5/3$.  Only the 
shocked ISM  is considered in the ST model.  Therefore, the shocked ISM mass
must significantly outweigh the mass of the material
ejected by the supernova (SN), usually at an age of a few thousand
years.  The governing parameters of this phase of SNR evolution are the
initial energy of the blast $E$, the density of the ambient medium $\rho_0$, 
and
the time since the explosion $t_0$. These parameters can be easily arranged to 
form a unit of length, related to the shock radius by
$r_s = 1.15 \left(Et_0^2/\rho_0\right)^{1/5}$.
These three initial parameters completely determine the dynamics of a Sedov
model. For example, we might choose an initial energy of $10^{51}$ ergs,
an upstream hydrogen number density $n_0 = 1$ cm$^{-3}$, and a time 
$t_0$ of $10^{11}$ seconds, yielding a radius of $2.5 \times 10^{19}$ cm. 
We now know the shock speed, $v_s=0.4r_s/t_0$, and also the postshock values 
of pressure, 
density, and velocity through the Rankine-Hugoniot shock conditions. The 
self-similar analytical solution (Sedov 1946; Taylor 1950) gives
the pressure, density and velocity profiles behind the shock, and
it allows us to find pressure and density variations as a function of time
for each fluid element. This information is
necessary for determination of the ionization state for each ion species.

Hamilton et al. (1983) showed that the shape of the X-ray spectrum of
a spatially-integrated Sedov model does not depend on all three parameters
mentioned above, but only on shock velocity $v_s$ (or alternatively
shock temperature 
$T_s = 1.37 \times 10^7 (v_s/1000 \kms)^2~{\rm K} = 
1.18 (v_s/1000 \kms)^2$ keV) and parameter $\eta=n_0^2E$. (We have
assumed cosmic-abundance plasma, where electrons are
provided by fully ionized H and He, with a negligible contribution from 
heavier elements.
The mean mass per particle $\mu$ is then equal to 0.6 in units of $m_p$.)
In order to simplify comparison with other NEI models,
we use ionization timescale $\tau_0$ instead of $\eta$, where we define 
$\tau_0$ as the product of the postshock electron density $n_{es}$ and the 
remnant's age $t_0$. Note that for cosmic-abundance plasma and the strong 
shock Rankine-Hugoniot jump conditions, $n_{es} = 4.8 n_0$. From equation
(4b) in Hamilton et al. (1983), ionization
timescale $\tau_0$ is equal to $6.03 \times 10^{11} \eta_{51}^{1/3}T_7^{-5/6}$
cm$^{-3}$ s, where $\eta_{51} = \eta/10^{51}$ ergs cm$^{-6}$ and 
$T_7=T_s/10^7$ K.
X-ray spectra depend also on electron temperature $T_e$ which is
discussed next.

\begin{figure}
\includegraphics[bb=0 0 504 360,width=3in,viewport=50 0 504 360,clip]{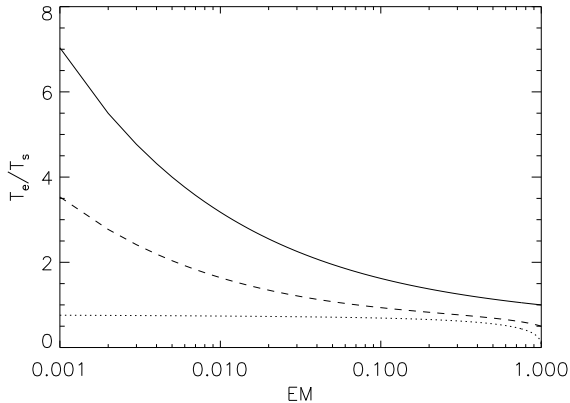}
%\includegraphics[bb=0 0 504 360,width=2.5in]{tedistr.eps}
%\plotone{tedistr.eps}

\caption{Electron temperature $T_e$ vs emission measure $EM$ (both normalized
to 1 at the shock), for Sedov model with 
$T_s = 3 \times 10^7$~K, $\tau_0 = 5 \times 10^{11}$ cm$^{-3}$ s, and with
full ion-electron equipartition ($T_e=T$; {\it solid curve}), no 
collisionless heating of electron at the shock 
({\it dotted curve}), and partial electron heating at the shock with the
postshock electron temperature $T_{es} = 0.5T_s = 1.5 \times 10^7$~K 
({\it dashed curve}). 
}
\label{electemp}
\end{figure}

\begin{figure}
\includegraphics[bb=0 0 481 566,width=3in,totalheight=0.35\textheight,viewport=40 0 481 566,clip]{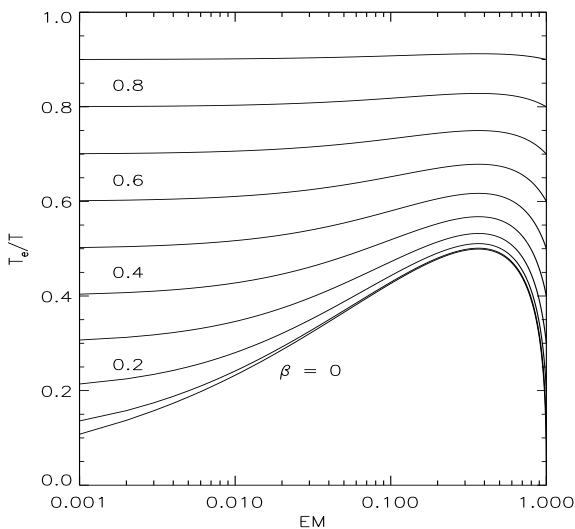}
%\includegraphics[bb=0 0 481 566,width=2.5in]{tedistrn.eps}
%\plotone{tedistrn.eps}

\caption{Ratio of electron temperature $T_e$ to the mean gas temperature $T$
vs emission measure $EM$, for the Sedov model with the same 
$T_s$ and  $\tau_0$ as in Figure~1, but with various amounts of collisionless
heating ($\beta$ ranges from 0.0 at the bottom to 0.9 at the top).
}
\label{electempn}
\end{figure}

\subsection{Electron Temperature}

The mean gas temperature $T$ in Sedov models is plotted in 
Figure~\ref{electemp} as a function of the normalized emission measure
$ EM(r) = \int_0^r n_e n_p dr/\int_0^{r_s} n_e n_p dr $. 
(Because we assumed normal abundances, with a cosmic H/He ratio and 
a negligible contribution to electron density from elements heavier than He,
models and calculations discussed in this work are not applicable to 
heavy-element plasmas where these assumptions break down.)
$T$ is equal to
the shock temperature $T_s$ at the shock ($EM=1$), and increases toward the
remnant's interior. But for calculations of X-ray spectra we need to know
the
electron temperature $T_e$, which in general is not equal to either the
mean temperature 
$T$ or the ion temperature $T_i$. These three temperature are equal only if
electrons and ions are equilibrated at the shock front through plasma 
instabilities generated within the collisionless plasma. 
Because kinetic energy is 
initially transferred to thermal motions of ions in collisionless shocks, full
equilibration demands the presence of a very efficient mechanism for transfer
of energy from ions to electrons in collisionless shocks, at all shock Mach
numbers. This is unlikely based on available observational and theoretical 
evidence. 
For example, observations of ultaviolet (UV) lines from a fast nonradiative 
shock in SN 1006 
(Laming et al. 1996) suggest that there is little energy transfer from ions 
to electrons in
shocks where neutral H is present ahead of the shock
front. In an extreme case of a very inefficient heating of electrons at the
shock, the postshock electron temperature $T_{es}$ is much less than the
postshock ion temperature $T_{is}$, and the ratio $T_{es}/T_{is}$ (or
$T_{es}/T_{s}$) is close to 0. Electrons are then heated mostly in Coulomb 
collisions with ions in the hot shocked plasma downstream of the shock front. 
The most general situation, most likely to happen in SNRs shocks, is that
electrons are partially heated in collisionless shocks and additional Coulomb
heating occurs downstream of the shock.

The extent of electron heating in shocks can be parametrized by the
temperature ratio $\beta \equiv T_{es}/T_s$, which can vary from nearly $0$
in shocks with negligible collisionless heating to $1$ when this heating is
very efficient. In general, $\beta$ might depend on the shock Mach number
and other shock parameters, possibly decreasing with increasing shock Mach
number (Bykov 
\&\ Uvarov 1999). This inverse correlation with the shock Mach number is
supported by most recent observational evidence based on optical
observations of nonradiative shocks (Ghavamian 1999; Ghavamian et al. 2000).
Because of the poor understanding of electron heating in collisionless shocks,
which may depend not only on the shock Mach number but also on other shock 
parameters such as the magnetic field orientation, we neglect dependence
of $\beta$ on shock parameters, i. e., we assume that $\beta$ does not vary 
during the SNR evolution. The validity of this assumption is questionable, and
should be examined in more detail once good quality spatially-resolved X-ray 
spectra of SNRs become available. Once a value of $\beta$ is chosen, we can
calculate the electron temperature $T_e$ in Sedov models using the procedure
described by Borkowski, Sarazin, \& Blondin (1994), which takes into account 
Coulomb heating
in the shocked plasma. (The formula ($D2$) in this paper contains 
a typographical error; the second
term on the right hand side of this expression should be replaced by 
$2.5\ln( (1+\beta^{1/2})/(1-\beta^{-1/2}) ) - 5\beta^{1/2}(\beta+3)/3$.)
The electron
temperature $T_e$ of a fluid element depends on its location within the
remnant, on $T_{es}$ and $T_s$ (hence on $\beta$), and on
the SNR ionization timescale $\tau_0$.

The extreme case of inefficient electron heating at the shock is shown in
Figure~\ref{electemp} for a Sedov model with $T_s = 3 \times 10^7$~K and 
$T_{es} = 0$~K ($\beta=0$), and $\tau_0 = 5 \times 10^{11}$ cm$^{-3}$ s. 
The electron temperature rises from 0 at the shock ($EM=1$) to an asymptotic
finite value at the SNR center ($EM=0$), the behaviour already noted 
by Itoh (1978, 1979) and Gronenschild \& Mewe (1979) in their  numerical 
simulations and discussed in more detail by Cox \& Anderson (1982) and 
Hamilton et al. (1983).
This is in contrast to very high electron temperatures in the interior
of models with efficient heating ($T_{es}=T_s$ or $\beta=1$). It is obvious 
that models with inefficient electron heating differ greatly from models with 
efficient heating. Another solution, with intermediate $\beta = 0.5$ 
($T_{es}=0.5T_s$) and the same $\tau_0 = 5 \times 10^{11}$ cm$^{-3}$ s, is 
also shown in Figure~\ref{electemp}. 

The relative importance of Coulomb heating and collisionless heating
in the Sedov model just discussed can be inferred from Figure~\ref{electempn},
where we plot the ratio of electron temperature $T_e$ to the mean temperature
$T$ for a number of different values of parameter $\beta$. For large $\beta$,
the curves are nearly straight parallel lines, with relatively unimportant
Coulomb heating. The electron temperature profiles in these models are nearly
the same as in the Sedov model with $\beta=1$, except for the overall 
temperature scaling which is proportional to $\beta$. At low $\beta$, the
Coulomb heating dominates, resulting in highly curved lines. The efficiency
of the Coulomb heating increases with increasing $n\Delta t/T^{3/2}$ 
(Itoh 1978), where $n$ is the total
particle density of a fluid element, $\Delta t$ is the time elapsed since this
element was shocked by the blast wave, and $T$ is the mean temperature of this
fluid element. At the blast wave and in the remnant's interior, 
$n\Delta t/T^{3/2}$ is small, so that the Coulomb heating is less efficient 
there than inside the swept shell. This is why the curves in 
Figure~\ref{electempn}
attain a maximum at $EM \sim 0.5$ and drop to 0 for $EM=0$ and~1. Note that
because of the scaling just mentioned, these curves are identical for all
Sedov models with the same value of $\tau_0/T_s^{3/2}$.

\begin{figure}
\includegraphics[bb=0 0 481 566,width=3in,totalheight=0.4\textheight,viewport=40 0 481 566,clip]{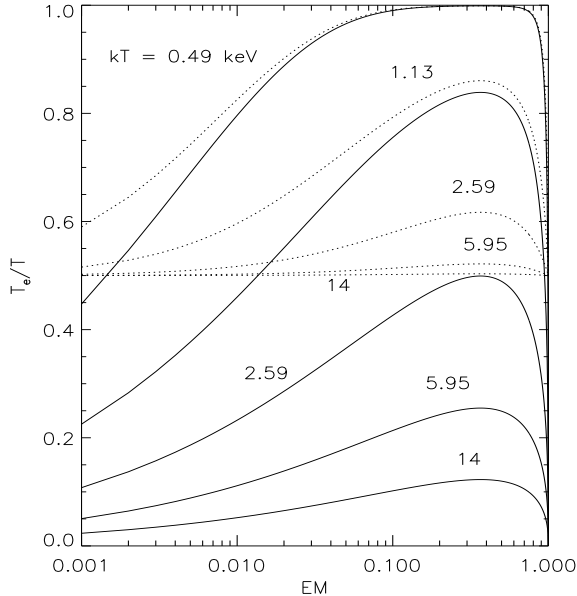}
%\plotone{tedistreta.eps}

\caption{Ratio of electron temperature $T_e$ to the mean gas temperature $T$
vs emission measure $EM$, for the Sedov model with 
$\eta = 8.88 \times 10^{51}$ ergs cm$^{-6}$,
at five different times (increasing from
bottom to top). Curves are labeled by the postshock temperature. Two cases 
with $\beta = 0$ ({\it solid curves}) and 
$\beta = 0.5$ ({\it dotted curves}) are shown.
 }
\label{electempeta}
\end{figure}

\begin{figure}
\includegraphics[bb=0 0 504 360,width=3in,totalheight=0.3\textheight,viewport=20 0 504 360,clip]{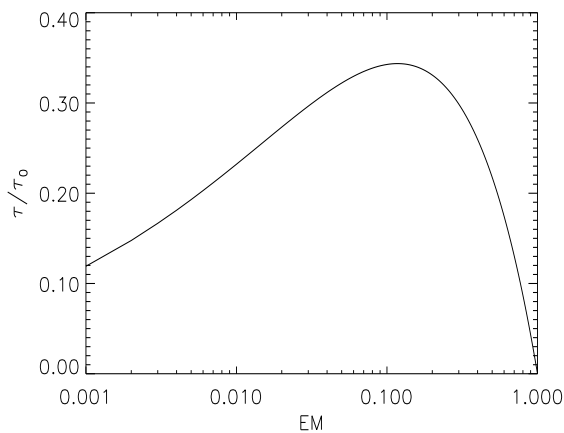}
%\plotone{itdistr.eps}

\caption{Normalized ionization timescale $\tau$ vs normalized emission measure
$EM$ for Sedov models.}
\label{ionizdistr}
\end{figure}

The temperature profiles presented in Figures~\ref{electemp} 
and~\ref{electempn} are merely examples, with a fixed value of 
$\tau_0/T_s^{3/2}$, which sample a small range of
the possible parameter space $( \beta,\ \tau_0/T_s^{3/2})$. 
The effects of
varying $\tau_0/T_s^{3/2}$ are shown in Figure~\ref{electempeta} for 
the same parameter $\eta$ as in Figures~\ref{electemp} and~\ref{electempn},
but at different times in the remnant's evolution (1/4, 1/2, 1, 2, and 4 of
$5 \times 10^{11}$ cm$^{-3}$ s). The requirement of constant $\eta$
demands that $T_s$ is equal to
$3 \times 10^7~{\rm K} / (n_0t/5 \times 10^{11}~{\rm cm}^{-3}~{\rm s})^{6/5}$.
Two cases of $\beta=0$ and 0.5 are shown. At early times, the fast 
collisionless heating in a model with $\beta=0.5$ dominates the Coulomb 
heating, leading to very different electron temperature profiles than for
$\beta=0$. This is the case discussed until now. The temperature profiles
become similar at later stages in the remnant's evolution, depending rather
weakly on $\beta$, because of the increased role of the Coulomb energy 
transfer. At $\tau_0 = 20 \times 10^{11}$ cm$^{-3}$ s (curves at the top
of Figure~\ref{electempeta}), noticeable differences are seen only in the  
remnant's interior, where densities are low and temperatures are high,
leading to a long electron-ion equlibration time scale. Note again that
these temperature profiles are identical for all Sedov models with the same 
values of $\tau_0/T_s^{3/2}$ and $\beta$.

\subsection{Ionization Calculations}

The ionization state of each fluid element depends on its ionization
timescale $\tau = \int_{t_s}^{t_0}n_e d\,t$, where $t_s$ is the time when
this fluid element was shocked, and on how the electron temperature 
$T_e$ varied
as a function of its ionization time scale, from 0 to $\tau$, during the 
remnant's evolution. The normalized ionization timescale $\tau/\tau_0$
is plotted in Figure~\ref{ionizdistr}. It starts at zero at the shock 
($EM=1$), then increases linearly in the swept SNR shell, but in
the remnant's interior it is again small because of low gas densities there.
Note that the distribution of $\tau$ is quite broad, and is poorly
represented by a $\delta$ function (the approximation involved in
using a single-ionization-timescale NEI model).
The most highly ionized atomic species are usually found near the maximum of
$\tau$, while the least ionized material is found just behind the shock and 
in the remnant's interior. We determined the electron temperature as a 
function 
of $\tau$ for each fluid element following the procedure described above.

We tracked the ionic states in each fluid element by solving the 
time-dependent ionization equations for each abundant element (C, N, O, Ne,
Mg, Si, S, Ca, Fe, and Ni). We used an eigenfunction method 
to solve these equations efficiently (Masai 1984, Hughes \&\ Helfand
1985, Kaastra \&\ Jansen 1993, Borkowski et al. 1994).
Physical processes included in these calculations
are collisional ionizations by electrons, both direct and involving 
excitations followed by autoionizations, and radiative and dielectronic
recombinations. We used collisional ionization rates from Arnaud \&\ Raymond 
(1992) for Fe, and from Lennon et al.~(1987) for other elements. We also
updated recombination rates in the Hamilton \&\ Sarazin (1984) X-ray code.
Radiative recombination rates were taken from P\'{e}quignot, Petitjean,
\&\ Boisson (1991), and rates for dielectronic recombination 
towards He-like, Li-like, and Be-like ions from Hahn (1993), and Teng, Xu,
\&\ Zhang (1994a,b). Recombination rates for Fe ions were taken 
from Arnaud \&\ Raymond (1992). Except for H and He, all elements are
assumed to be neutral initially, a good approximation under most
circumstances because of the insensitivity of X-ray spectra to the detailed
ionization state of preshock plasma.

\subsection{Spectral Code}

We used an updated version of the X-ray code written by Hamilton \&\ Sarazin
(1984) to calculate X-ray spectra. We updated collision strengths for H-like
ions using fits provided by Itikawa et al.~(1985) for O$^{+7}$ and by 
Callaway (1994) for other H-like ions. Collision strengths for He-like ions
were taken from Sampson et al.~(1983), Keenan et al.~(1987), and 
Kato \&\ Nakazaki (1989). We also updated atomic data for fluorescent
Fe K$\alpha$ lines produced by inner-shell ionization.
K-shell collisional ionization rates by
electrons for all Fe ions were calculated according to Beigman, Shevelko,
\&\ Tawara (1996), and then combined with the fluorescence yields
of Kaastra \&\ Mewe (1993) to obtain strengths of fluorescent K$\alpha$ lines
for Fe$^{+0}$ -- Fe$^{+21}$. The fluorescent yield for Fe$^{+22}$ was taken
from theoretical calculations of
transition probabilities and autoionization rates by Seely, Feldman, 
\&\ Safronova (1986). 
Energies, transition probabilities, and excitation rates by electrons for 
Fe L-shell transitions for 
Fe$^{+16}$ -- Fe$^{+23}$ were kindly provided to us by Duane Liedahl 
(see Liedahl, Osterheld, \&\ Goldstein 1995 and Mewe, Kaastra, \&\ Liedahl 
1995 for the description of these theoretical calculations and their 
implementation into the equilibrium ionization thermal MEKAL model).

\subsection{Sedov Models in XSPEC}

Modeling X-ray spectra of SNRs within the framework of Sedov models demands 
an efficient interface between observations and the calculated model spectra.
Such an interface is provided by the well-known and widely-used software 
package XSPEC (Arnaud 1996). We wrote efficient FORTRAN programs for 
calculations of Sedov X-ray spectra, using methods and techniques just
described, and constructed Sedov models in XSPEC v9 and v10. (The most recent
version of these models was implemented by Keith Arnaud into XSPEC v11.) 
The model parameters
are: post-shock temperature $T_s$, post-shock electron temperature $T_e$, 
ionization timescale $\tau_0$, and elemental abundances. Following standard 
practice used with thermal models in XSPEC, there are two versions of Sedov
models, where heavy-element abundances may be varied individually or together.

Calculations of Sedov models are computationally intensive, resulting in very
long computational times when X-ray data sets are fitted with these models.
An even more troublesome aspect is the lack of smooth convergence when fitting
with Sedov models. Keith Arnaud solved this problem by modifying a routine
in the FTOOLS software package which was used to tabulate MEKAL model. With
this routine we 
can now tabulate spatially-integrated Sedov models in a FITS file, which can
then be used in conjunction with a standard XSPEC model ``atable'' to make
fits to the data. This procedure is necessary for the robust performance of
the Sedov models in XSPEC.

\subsection{Calculated Spectra: Effects of Collisionless Electron Heating} \label{calcspectra}

The flexible computer interface provided by the XSPEC package allows us to 
easily generate Sedov X-ray spectra in the whole parameter space relevant for
SNRs, in the energy range from 0.1 keV to 30 keV, and to simulate observations
using spectral response matrices and effective areas of modern X-ray 
satellites. There are clearly a number of interesting 
issues associated the use of Sedov models, a lot of them discussed in detail
by Hamilton et al.~(1983), which can be efficiently examined within the XSPEC
package. Here we restrict ourselves to the effects of collisionless electron
heating on X-ray spectra of SNRs.  

\begin{figure}
\includegraphics[bb=0 0 481 566,totalheight=0.6\textheight,width=3in,viewport=50 0 481 566,clip]{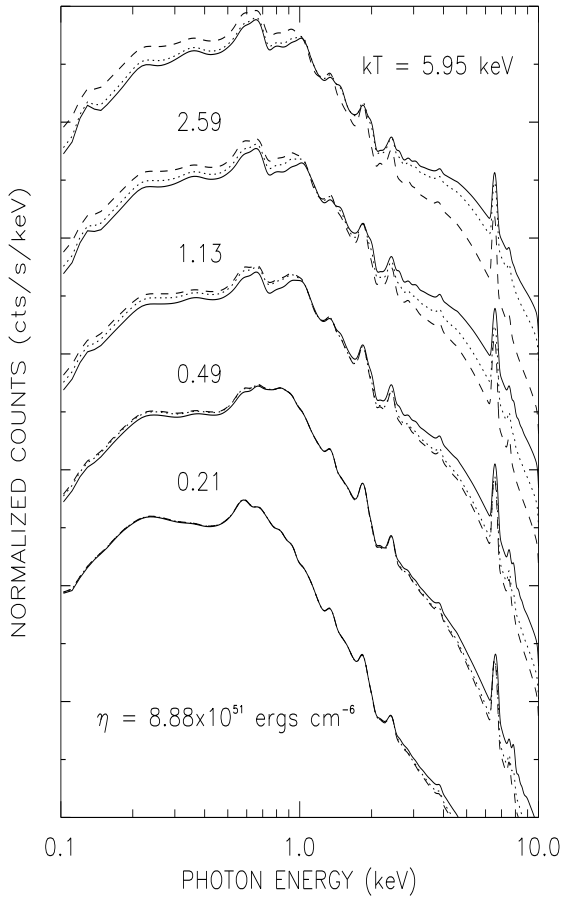}
%\plotone{spectr259.eps}

\caption{X-ray spectra for Sedov models with $\eta = 8.88 \times 10^{51}$ ergs
cm$^{-6}$ and with various amounts of collisionless heating:
full electron-ion equipartition ({\it solid curve}), no 
collisionless heating of electrons at the shock ({\it dashed curve}),
and partial electron heating at the shock with the
postshock electron temperature $T_{es} = 0.5T_s$ ({\it dotted curve}). 
Models are displaced from each other by a factor of 10 in the vertical 
direction and they are labeled by the postshock temperature (in keV).
}
\label{spectr259}
\end{figure}

X-ray spectra for Sedov models with $\eta=8.88 \times 10^{51}$ ergs cm$^{-3}$
are shown in Figure~\ref{spectr259} at different times during the remnant
evolution, and at each time for three values of $\beta$: 0.0, 0.5, and 1.0.
All spectra were folded through the pre-launch instrumental response of the 
back-illuminated ACIS CCD detector (S3) onboard the {\it Chandra} Observatory, 
and 
include variations of the telescope's effective area with the photon energy.
The second set of spectra from the top of Figure~\ref{spectr259} corresponds
to models whose electron temperature profiles are shown in 
Figure~\ref{electemp}. These spectra differ substantially from each other,
and become harder with the increasing amount of collisionless heating. 
Particularly large differences are seen at high energies, reflecting 
large differences in temperature profiles seen in Figure~\ref{electemp}.
It should be obvious that, in general, {\it it is necessary to consider
not only the extremes of very efficient and completely inefficient 
electron heating, but also models with partial electron heating at the shock}.

Differences between X-ray spectra of Sedov models with different amounts of 
collisionless heating become smaller as the remnant ages, because of the
increased role of the Coulomb heating in older remnants without full 
electron-ion equilibration at the shock front (see Figure~\ref{electempeta}).
This is clearly seen in Figure~\ref{spectr259}, where there is hardly any
difference between the coolest models with and without collisionless heating.
Most of the X-ray emitting plasma in these models is actually not very far from
coronal equilibrium ionization.

\section{INTERPRETATION OF X-RAY SPECTRA}

X-ray and UV spectra of stellar coronae showed the presence of
multi-temperature
hot plasmas, which led researchers to consider temperature distribution
functions in their efforts to understand the observed coronal spectra. This
approach proved quite useful in analysis of X-ray spectra of stellar coronae.
It would be advantageous to have a similar framework available for SNRs, so
we generalized the idea of the temperature distribution function used for
plasmas in ionization equlibrium to NEI Sedov models.

\subsection{Temperature and Ionization-Timescale Distribution Functions}

\begin{figure}
\includegraphics[bb=0 0 481 566,width=3in,totalheight=0.35\textheight,viewport=50 0 481 566,clip]{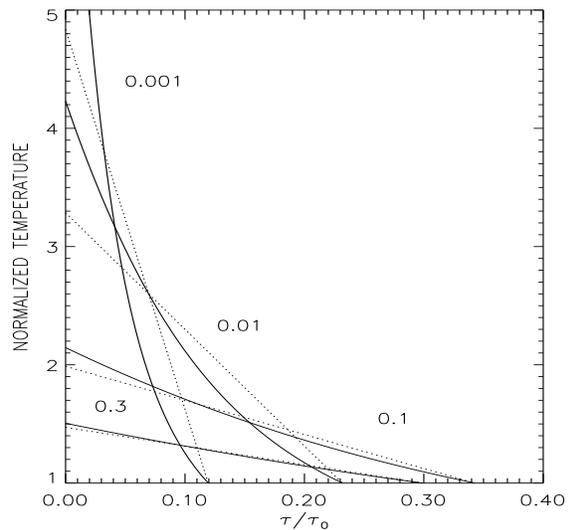}
%\includegraphics[bb=0 0 481 566,width=3in]{addfig.eps}
%\plotone{addfig.eps}

\caption{Temperature of fluid elements in the Sedov model, normalized to their
final values, vs normalized ionization timescale $\tau/\tau_0$.
The temperature history is shown for 4 different fluid
elements ({\it solid curves}), labeled by their normalized emission measure 
$EM$. A linear approximation discussed in the text is also shown 
({\it dotted curves}).} 

\label{addfig}
\end{figure}

\begin{figure}
\includegraphics[bb=0 0 481 566,width=3in,totalheight=0.35\textheight,viewport=50 0 481 566,clip]{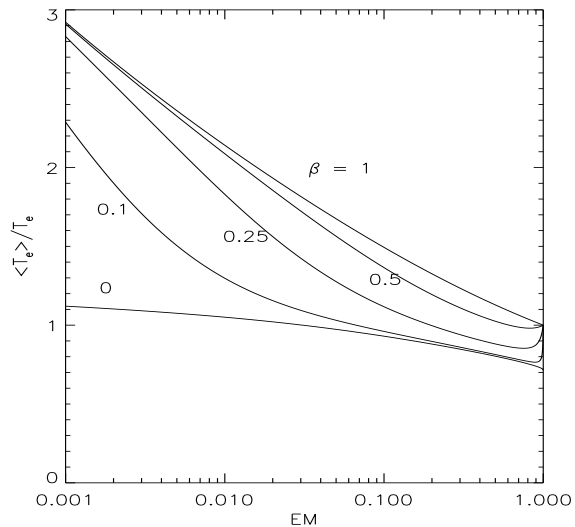}
%\plotone{teavrg.eps}

\caption{Ratio of ionization timescale-averaged electron temperature $\langle T_e \rangle$ to 
final electron $T_e$ vs emission measure $EM$ for Sedov models shown in
Figures~1 and 2, and with various amounts of collisionless heating. Curves
are labeled by $\beta$. }
\label{avrgte}
\end{figure}

The spatially-integrated X-ray spectra for Sedov models can be succesfully
interpreted in terms of the following three distributions: the electron
temperature, ionization timescale, and ionization timescale-averaged 
electron temperature distribution functions. We have already shown the 
ionization timescale distribution in Figure~\ref{ionizdistr}, and examples of the electron 
temperature distribution in Figure~\ref{electemp}. These two distributions do not contain 
enough
information to determine ionic states, because of the adiabatic expansion
of the gas following its passage through the blast wave. This expansion 
results in a decreasing mean temperature for each fluid element.
This is demonstrated in Figure \ref{addfig}, where we plot temperature
variations for several fluid elements in the Sedov model. Particularly
large variations are seen for fluid elements which were shocked early in the
remnant's evolution. 
Another
reason for electron temperature variations is energy transfer from ions to
electrons in models with unequal ion and electron temperature. In principle 
a detailed account of the electron temperature variations is needed
for each fluid element, just as we did for Sedov models, but a simpler
approach in terms of an average temperature provides a better insight while 
maintaining good accuracy. A suitable average 
electron temperature is the ionization timescale-averaged temperature 
$\langle T_e \rangle$, defined as 
$\int_{t_s}^{t_0} T_e(t) n_e(t) dt / \tau $. In Figure~\ref{avrgte}, 
we plot the 
ratio $\langle T_e \rangle  /T_e$ for Sedov models with 
$T_s = 3 \times 10^7$~K, $\tau_0 = 5 \times 10^{11}$ cm$^{-3}$ s, and various
amounts of collisionless electron heating. For equal ion and electron
temperatures this
ratio is significantly greater than unity in the remnant's interior 
because of the adiabatic expansion. In the opposite limit
of no collisionless heating, $\langle T_e \rangle$ is slightly less than $T_e$ 
in the postshock region, because the efficient transfer of 
energy from ions to electrons more than compensates for adiabatic expansion.
However, even in this case the temperature ratio becomes larger than unity in 
the remnant's interior, demonstrating the effectiveness of the adiabatic
expansion. After passing through the shock front, electrons in a fluid
element are heated by
Coulomb collisions with ions, but because of the adiabatic expansion the
efficiency of Coulomb heating decreases, and the electron temperature attains
a maximum and then starts to decline with time. Even modest amounts of 
collisionless heating affect the ionization timescale averaged distribution
function in the remnant's interior and at the shock front 
(Fig.~\ref{avrgte}), where the ionization timescale is low and the Coulomb
heating is least effective (Fig.~\ref{electempn}). As the collisionless
heating becomes more important, the ratio $\langle T_e \rangle  /T_e$ increases
until it reaches its upper limit in fully equilibrated models. 

\begin{figure}
\includegraphics[bb=0 0 481 566,width=3in,totalheight=0.6\textheight,viewport=50 0 481 566,clip]{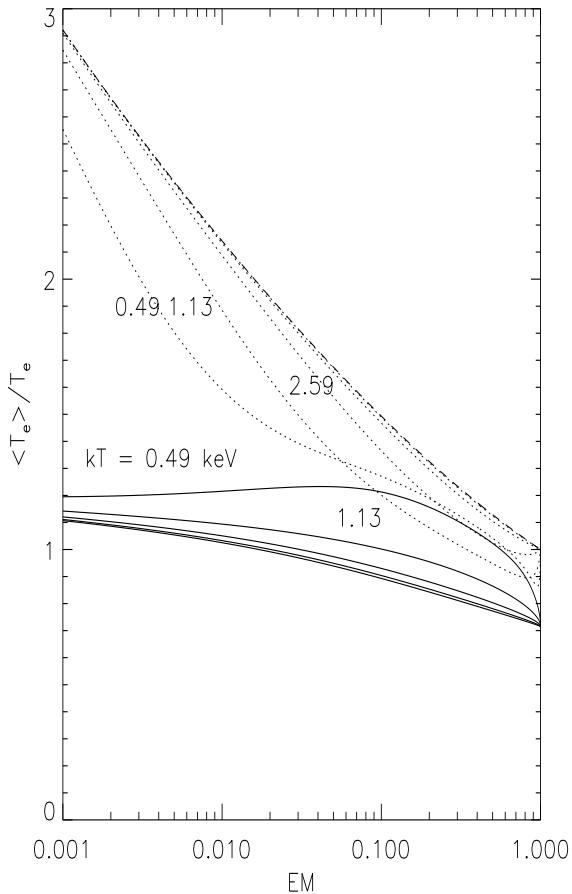}
%\plotone{teavrgeta.eps}

\caption{Ratio of ionization timescale-averaged electron temperature $\langle T_e \rangle$ to 
final electron $T_e$ vs emission measure $EM$ for Sedov models shown in
Figure~5, with $\eta = 8.88 \times 10^{51}$ ergs. Two cases 
with $\beta = 0$ ({\it solid curves}) and 
$\beta = 0.5$ ({\it dotted curves}) are shown. Curves are labeled by the shock
temperature. {\it Dashed curve} corresponds to model with full electron-ion
equilibration.}
\label{avrgteeta}
\end{figure}

The temperature profiles presented in Figure~\ref{avrgte} are again 
merely examples, with a fixed value of $\tau_0/T_s^{3/2}$, which sample a 
small range of the possible parameter space $(\beta, \ \tau_0/T_s^{3/2})$. 
The effects of varying $\tau_0/T_s^{3/2}$ are shown in 
Figure~\ref{avrgteeta}, for Sedov models with electron temperature distribution
functions shown in Figure~\ref{electempeta}. The ionization timescale
averaged distributions differ from each other, reflecting the relative
importance of adiabatic cooling and Coulomb heating at various locations within
the remnant. However, they are all bounded from below by models with low
efficiency of Coulomb heating and no collisionless heating, and from above by
models with equal ion and electron temperatures.

The two temperatures, the electron temperature $T_e$ and the average 
electron temperature $\langle T_e \rangle$, may be used to obtain
an approximate functional
relationship of electron temperature with ionization timescale for each fluid
element. A linear relationship is simple and sufficient, starting with 
temperature
$2\langle T_e \rangle - T_e$ at the shock (ionization timescale equal to 0), 
and ending with $T_e$ at 
the current local ionization timescale $\tau$.  We show this linear
approximation in Figure \ref{addfig} for a model with equal ion and electron
temperatures. The agreement with the exact Sedov solution is excellent for
the bulk of the X-ray emitting material, although it becomes somewhat less 
accurate in the remnant's interior where temperature variations are more 
pronounced. This approximate description
of electron temperature variations allows us to calculate ionization
fractions and then X-ray spectra, and
compare them with the exact X-ray calculations for Sedov models described in
\S~\ref{calcspectra}. We find excellent agreement between the two sets of 
calculations for models with equal ion and electron temperatures,
with typical differences at most of order of few percent (for cosmic abundance
Sedov models at the ASCA or {\it Chandra} CCD spectral resolution). The 
agreement is slightly worse for models without any collisionless heating, 
but the differences are rarely at the $\sim 10$\%\ level and they are
limited to a few line complexes at lower energies. 
Because such differences are smaller than current uncertainties in atomic 
data, we did not find it necessary to develop a more accurate approximation
to describe the evolution of electron temperature in a fluid element.
The current linear approximation
is clearly sufficient for calculations of X-ray
emission from the Sedov blastwave. While it offers negligible
computational advantages over exact calculatons for Sedov models, we find it 
useful for interpretation of X-ray spectra. 
This approximate description may
also be very convenient in more complex situations, which might require 
multi-dimensional hydrodynamical simulations.

X-ray spectra of SNRs are 
apparently described very well by the electron temperature, ionization 
timescale, and ionization timescale averaged electron temperature distribution
functions. High-quality X-ray data may allow us to determine these
distribution functions without reliance on detailed hydrodynamical 
calculations. This is analogous to determinations of differential emission
measure in stellar coronae, without reliance on coronal models. The experience
gained in analysis of stellar coronae suggests that such an empirical approach 
should be useful in analysis of X-ray spectra of SNRs. But instead of one
temperature distribution function needed in a stellar corona, three 
distribution functions are required to describe NEI X-ray spectra of SNRs. 
A purely empirical 
determination of these distribution functions might then be a difficult task,
so we expect that its combination with hydrodynamical modeling will prove
most useful in the future.

\subsection{Constant Temperature, Single Ionization Timescale Models}

\begin{figure}
\includegraphics[bb=0 0 481 566,width=3in,totalheight=0.6\textheight,viewport=50 0 481 566,clip]{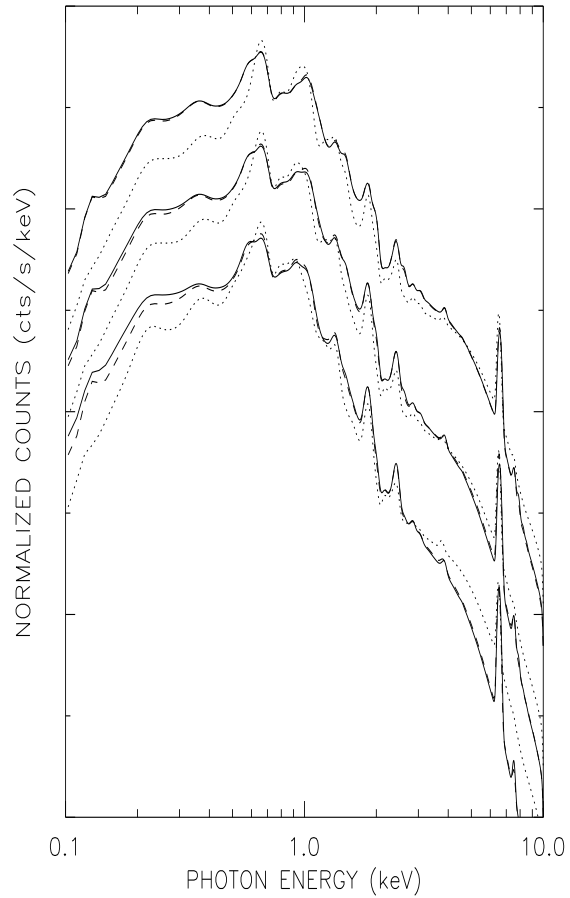}
%\plotone{shckfits.eps}

\caption{Fits to X-ray spectra of Sedov models of Figure~3, for clarity 
displaced 
from each other by factor of 10 and shown by {\it solid curves} 
(top -- $T_{es}=T_s$, middle -- $T_{es}=0.5T_s$, bottom -- $T_{es}=0$). 
A constant temperature, single ionization timescale
model ({\it dotted curves}) provides a poor fit when compared with 
plane-parallel shock ({\it dashed curves}).}
\label{figsedt259t5e11}
\end{figure}

The most common NEI model in use today is the constant temperature, single
ionization timescale model. In terms of the distribution functions just
discussed, electron temperature and average electron temperature are assumed
constant, and the ionization distribution shown in Figure~\ref{ionizdistr} 
is approximated
by a delta function. As already noted by researchers (e.~g., Masai 1994), the
latter approximation is particularly troublesome. If it is used to describe
a plane-parallel shock or a Sedov model, prominent lines produced by strongly
underionized plasma immediately behind the shock will be missing from the
model spectrum. This is demonstrated in Figure~\ref{figsedt259t5e11}, where 
we again show the calculated X-ray spectra for Sedov model with 
solar abundances, and with the same parameters as in 
Figures~\ref{electemp} and~\ref{avrgte}
($T_s=3 \times 10^7~{\rm K} = 2.59~{\rm keV}, \tau= 5 \times 10^{11}$ cm$^{-3}$ s).
We fitted a simple NEI model just described to these spectra, using the XSPEC
software package, and the best fits are shown in Figure~\ref{figsedt259t5e11}. 
The very poor match 
between the calculated spectra and the best fit model spectra 
demonstrates that the model under consideration is not adequate to represent
X-ray spectra of SNRs. {\it The current practice of using the constant
temperature, single ionization timescale NEI model to represent X-ray spectra
of SNRs is not valid in general.} This simple NEI model may be used 
judiciously 
under special circumstances, e.~g., for an isolated dense cloud of gas overrun 
by a shock long time ago or for SN ejecta in old remnants which were 
thermalized early in the remnant's evolution. However, these are exceptions
to the rule because X-ray emission in SNRs is generally produced behind 
shocks, much as in Sedov models. 

\subsection{Plane-Parallel Shock Models}

The failure of a constant temperature, single ionization timescale model is
caused by an improper approximation of the ionization timescale distribution
(Fig.~\ref{ionizdistr}) by the delta function. A broad distribution such as seen in
Figure~\ref{ionizdistr} is needed in order to provide a good approximation to Sedov models.
A simple and obvious choice is a plane-parallel shock characterized by a
constant postshock electron temperature $T_e$ and by its ionization age 
$\tau_s = n_e t_s$, where $n_e$ is the postshock electron density and 
$t_s$ is the
shock age. This model differs from a single ionization timescale NEI model in 
its linear distribution of ionization timescale vs emission measure (at the 
shock the ionization timescale is 0, and it attains the maximum 
$\tau_s$ for the material
shocked earliest). Such a simple shock provides a much better approximation to
the Sedov models shown in Figure~\ref{figsedt259t5e11}, where we also plot the 
best fit shock model.
The difference between Sedov models and the shock model is noticeable only
at low energies, particularly for models with no collisionless electron 
heating. For equal ion and electron temperatures, final 
values of the free parameters of the fit give a very good
approximation to the mean electron temperature ($T_e = 3.26$ keV vs the 
emission-weighted electron temperature of $1.27T_s=3.29$ keV in the Sedov 
model) and the
mean ionization timescale ($\tau_s/2 = 9.88 \times 10^{10}$ cm$^{-3}$ s vs 
$\langle \tau \rangle = 0.202 \tau_0 = 1.01 \times 10^{11}$ cm$^{-3}$ s, where 
$\langle \tau \rangle$ is the emission-measure averaged 
ionization timescale in Sedov models). Reasonable agreement is also obtained
for the other two nonequipartition models with $T_{es}=0.5T_s=1.30$ keV and
$T_{es}=0$: shock temperatures 2.15 keV and 1.58 keV vs the emission-weighted
electron temperatures of $0.74T_s=1.91$ keV and $0.53T_s=1.37$ keV in Sedov
models, and $\tau_s/2=7.84 \times 10^{10}$ cm$^{-3}$ s,
$6.50 \times 10^{10}$ cm$^{-3}$ s vs $1.01 \times 10^{11}$ cm$^{-3}$ s in
Sedov models. This example 
demonstrates that plane-parallel shocks might serve as useful approximations to
Sedov models with high electron temperatures in the {\it Chandra} spectral 
range. From
the computational point of view, plane-parallel shocks with constant electron
temperature require little computational overhead when compared with the single
ionization timescale model (when ionic states are calculated with the
eigenfunction technique). But unlike the latter model, plane-parallel shocks 
are useful for fitting X-ray spectra in  a broad range of model parameters.

Plane-parallel shocks with constant electron temperatures just discussed 
constitute a subclass of more general plane shocks with partial electron
heating at a shock. Electrons are also heated by collisions with ions
downstream of the shock, just as in Sedov models. These shocks may be 
parametrized by the shock 
temperature $T_s$, the postshock electron temperature $T_{es}$, and the shock
ionization age $\tau_s$. If $T_s=T_{es}$, we arrive at constant-temperature 
shock models discussed above. The case with no collisionless heating at the
shock corresponds to $T_{es}=0$, where Coulomb collisions gradually
lead to increase in the electron temperature $T_e$ downstream of the shock, 
which becomes equal to $T_s$ in the limit of the infinite shock ionization age 
$\tau_s$, far from the shock front. For partial electron heating at the shock,
$0 < \beta = T_{es}/T_s < 1$, just like for Sedov models.

Multitemperature, plane-parallel shock models with unequal electron and ion 
temperatures at the shock should generally provide better fits to Sedov models.
We find that this is indeed true, in particular, we can obtain perfect fits
to Sedov models in Figure~\ref{figsedt259t5e11}. For $T_{es}=0.5T_s=1.30$ keV,
the best plane-parallel shock fit gives 
$T_s=5.57$ keV, $T_{es}=0.82$ keV, $\tau_s/2=1.08 \times 10^{11}$ cm$^{-3}$ s,
and emission weighted electron temperature $T_e=0.37T_s=2.04$ keV. For 
$T_{es} = 0$, we get 
$T_s=3.36$ keV, $T_{es}=0$,  $\tau_s/2=9.66 \times 10^{10}$ cm$^{-3}$ s, and
emission weighted electrom temperature $T_e=0.44T_s=1.46$ keV. It may be seen
that while $T_s$ and $T_{es}$ cannot be reliably determined by fitting shocks
to Sedov models of Figure~\ref{figsedt259t5e11}, the derived emission weighted 
$T_e$ and the SNR ionization age $\tau_0$ compare well with parameters of the
Sedov model. 

\begin{figure}
\includegraphics[bb=0 0 481 566,width=3in,totalheight=0.6\textheight,viewport=50 0 481 566,clip]{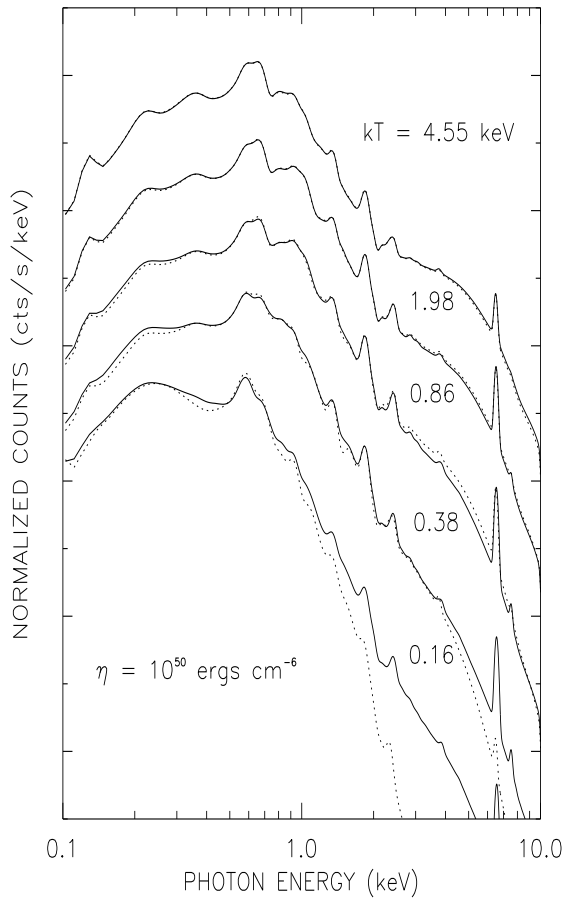}
%\plotone{shckfitseta50.eps}

\caption{X-ray spectra for Sedov models with $\eta = 10^{50}$ ergs cm$^{-6}$
and with equal ion and electron temperatures, for clarity displaced from
each other by a factor of 10 in the vertical direction. Curves are labeled
by the postshock temperature (in keV). The best fit shock models 
({\it dottted curves}) compare well with Sedov models ({\it solid curves}) 
at high temperatures, but they differ significantly at low temperatures.}

\label{shckfitseta50.eps}

\end{figure}

The excellent agreement between plane-parallel shocks and Sedov models at
high electron temperatures, demonstrated in Figure~\ref{figsedt259t5e11}, 
does not hold at low shock velocities. In Figure~\ref{shckfitseta50.eps},
we show X-ray spectra for an SNR with 
$\eta = n_0^2E = 10^{50}$ ergs cm$^{-3}$ and equal ion and electron
temperatures, at various times during its evolution.
The best fit spectra produced by 
a plane-parallel, constant electron temperature shock are also shown. The
fits are perfect early in the remnant's evolution, at high shock temperatures,
but they deteriorate as the blast wave decelerates. Particularly large
differences are seen at high energies, where the shock model
spectrum is too soft. This is not suprising because high energy emission in
Sedov models is produced at significantly higher temperatures than at the 
shock, as can be seen from the electron temperature distribution function
in Figure~\ref{electemp}. Constant temperature plane parallel shocks 
provide a poor approximation to this multitemperature distribution function,
and should not be generally used to fit X-ray spectra of low temperature 
Sedov models in the 0.1 -- 10 keV spectral range.

\subsection{Multitemperature NEI Models}

An inadequate representation of the distribution functions is the reason why
single ionization timescale NEI models in general, and plane-parallel shocks
at low shock speeds, fail to reproduce X-ray spectra of Sedov models. In the
latter, it is the presence of multitemperature plasma in Sedov models (see
Figures~\ref{electemp} and~\ref{avrgte}) which is responsible for this failure. It is possible to
construct two-temperature, three-temperature, and generally multi-zone 
``simple'' NEI models which
provide a good approximation to X-ray spectra of Sedov models, provided that
these models are consistent with the three distribution functions discussed
previously. We constructed such models succesfully, and found that at most
$\sim 10$ temperature zones are needed for Sedov models at low temperatures, 
while
just a couple might be sufficient at intermediate and high temperatures. While
such multitemperature NEI models might possibly be useful for computational
reasons, they have been designed as a method of successive approximation (more
temperature zones give a better model), and not by a physical simplicity.
For this reason we advocate the use of full Sedov models, and not
multi-zone models, in modeling of spatially-integrated X-ray spectra of SNRs, 
because
Sedov models represent a simple physical situation of a spherical blast wave
propagating into a uniform ambient medium. 
However, multi-zone models most likely will be very useful in 
spatially-resolved X-ray spectral modeling of SNRs.

\section{DETERMINATION OF MODEL PARAMETERS FROM X-RAY SPECTRA}

There are three parameters of Sedov models which may be determined from
spatially integrated X-ray spectra: postshock temperature $T_s$, 
postshock electron temperature $T_{es}$, and the ionization age $\tau_0$ 
(plus chemical abundances which we do not discuss here). It is generally
difficult to determine all three parameters, particularly when observations
are available only
in a limited X-ray spectral range or when deviations from the Sedov dynamics
are present. We will first consider two limiting cases of young and hot, 
and much older and cooler SNRs.

When young and hot remnants are observed in the 0.1 -- 10 keV spectral range,
such as provided by the {\it Chandra} CCDs, most of the emission
comes from the bulk of the X-ray emitting material, with emission measures
between 0.1 -- 1.0, located behind the blast
wave. Because of the low ionization timescales characteristic of young
remnants, the electron temperature distribution function is generally flat 
and has a similar shape for models with various amounts of electron 
heating in this range of emission measures (see Figures~\ref{electemp} 
and~\ref{electempn}). The ionization timescale averaged electron temperature
distributions are also similar (Figure~\ref{avrgte}), while the ionization
timescale distributions are of course identical and reasonably well 
approximated by a straight line (Figure~\ref{ionizdistr}). We have already
shown that under these conditions plane parallel shocks with constant electron
temperature provide an excellent approximation to Sedov models in the 
0.1 -- 10 keV energy range. This demonstrates that the observed
X-ray spectrum depends mostly on the mean electron temperature in the postshock
region and on the ionization timescale $\tau_0$. It is difficult to determine
$T_s$ and $T_{es}$ separately without additional information. Observations
at higher energies, which would probe interior regions where the temperature
distribution functions differ between the models (see Figure~\ref{electemp}),
are not likely to succeed because of expected severe deviations from the 
Sedov dynamics in the interior regions of young remnants.

It is possible to determine independently the shock speeds in young remnants
through high spectral resolution observations. The postshock radial velocity 
of the bulk of the X-ray emitting material is equal to $0.75 v_s$, which 
results in a velocity difference of $1.5 v_s$ between the front and back 
section of the shell at the remnant's center. With a sufficient spectral
resolution, Doppler-shifted spectral lines originating in these two shell 
sections could be separated, providing us with the missing information 
about the shock speed. This straightforward kinematic measurement is 
currently hard to accomplish because of the lack of X-ray instrumentation with 
a sufficiently high spectral resolution for spatially-extended sources such 
as SNRs. Such measurements would provide us with
the shock temperature, which in combination with the analysis described
above would allow for a complete determination of Sedov model parameters.
Note that the high spectral resolution would also provide  
additional electron temperature and ionization timescale diagnostics based on 
individual line ratios.

Additional information about postshock velocities and postshock electron
temperatures can also be independently found from optical and UV 
observations of nonradiative, Balmer-dominated shocks. The width of the broad 
H$\alpha$ component seen in these shocks is a direct measure of the postshock 
proton temperature. In combination with X-ray observations, this is enough
information to determine the shock speed in young remnants. In addition, 
UV and optical lines in such shocks are produced by electron impact excitation,
(although in the fastest shocks proton excitation might also be important),
and carefully chosen line ratios can be good diagnostics of postshock electron
temperatures. This is a poweful method for determining both temperatures from
optical observations alone,
although it is limited to shocks with a nonnegligible population of neutral
H atoms entering the shock front. But for such fast nonradiative shocks one 
can find a complete set of shock parameters from optical, UV, and X-ray 
observations.

In contrast to hot and young SNRs just discussed, low shock temperatures and
large ionization timescales in old remnants lead to efficient Coulomb heating
and ion-electron equilibration. Departures from equilibration are important
only in the immediate vicinity of the shock and in the remnant's interior
(e.g., see top of Figure~\ref{electempeta}). Spatially-integrated X-ray
spectra for models with the same postshock temperature $T_s$ but different
$T_{es}$ differ significantly only at high energies, as shown in 
Figure~\ref{spectr259} for a model with
$kT_s = 0.49$ keV and $\eta = 8.88 \times 10^{51}$ ergs cm$^{-6}$. The
postshock temperature $T_s$ and the ionization timescale $\tau_0$ 
(or parameter $\eta$) can be obviously found from the shape of the 
X-ray spectrum at low energies. But determination of the postshock electron 
temperature $T_{es}$ must rely on the high-energy X-ray spectrum.
The high-energy spectrum is produced by hot gas in the remnant's interior
which was shocked early in the evolution of the remnant when departures
from the Sedov dynamics could be significant. In this situation, it might
be difficult to distinguish between effects of collisionless electron
heating and effects of departures from the Sedov dynamics, particularly 
for remnants with low shock temperatures $T_s$ and large ionization 
timescales $\tau_0$. For such remnants, determination of both $T_s$ and 
$T_{es}$ is best accomplished through spatially-resolved X-ray, UV, and 
optical spectroscopy in the narrow region immediately behind the shock 
front (e. g., see Ghavamian 1999; Ghavamian et al. 2000).

\begin{figure}
\includegraphics[bb=0 0 481 566,width=3in,totalheight=0.6\textheight,viewport=50 0 481 566,clip]{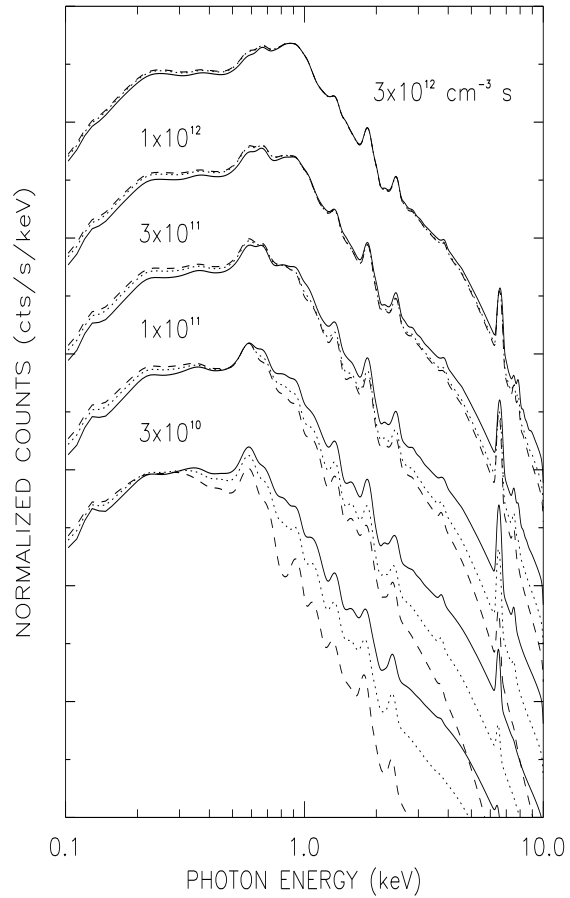}
%\plotone{spectr06.eps}

\caption{X-ray spectra for Sedov models with $T_s = 0.6$ keV 
and with various amounts of collisionless heating:
full electron-ion equipartition ({\it solid curve}), no 
collisionless heating of electrons at the shock ({\it dashed curve}),
and partial electron heating at the shock with 
$T_{es} = 0.5T_s = 0.3$ keV ({\it dotted curve}). 
Models are displaced from each other by a factor of 10 in the vertical 
direction and they are labeled by the ionization timescale $\tau_0$ 
(in cm$^{-3}$ s).}

\label{spectr06}

\end{figure}
The postshock electron temperature $T_{es}$ and the postshock temperature
$T_s$ can be simultaneously determined from spatially-integrated spectra
for remnants with intermediate shock temperatures and low ionization 
timescales $\tau_0$. We demonstrate this in Figure~\ref{spectr06}, where
we show calculated X-ray spectra for Sedov models with the postshock
temperature $T_s = 0.6$~keV, with three values of $T_{es}$ (0 keV,
0.3 keV, and 0.6 keV), and for $\tau_0$ ranging from 
$3 \times 10^{10}$ cm$^{-3}$ s to $3 \times 10^{12}$ cm$^{-3}$ s. (Again,
the calculated spectra were folded through the spectral response of the 
back-illuminated ACIS CCD detector.) Note that the plane-parallel shock 
model provides a poor approximation to these spectra because of the 
relatively low (0.6 keV) postshock temperature. X-ray spectra for
models with low values of $\tau_0$ show strong dependence on $T_{es}$,
particularly at high energies, because of the lack of ion-electron 
equilibration. For such remnants, a simultaneous determination of $T_s$
and $T_{es}$ from spatially-integrated X-ray spectra should be possible.
As the ionization timescale $\tau_0$ increases, Coulomb heating becomes
effective, and the effects of collisionless electron heating decrease
in importance. For large ionization timescales $\tau_0$, spectra in 
Figure~\ref{spectr06} depend little on $T_{es}$, and determination of
$T_{es}$ becomes difficult. 

\section{SUMMARY}

We discussed spatially-integrated X-ray spectra of SNRs based on Sedov 
dynamics. Their spectral shape depends on elemental abundances and on three 
parameters: postshock temperature $T_s$, postshock electron temperature 
$T_{es}$, and ionization timescale $\tau_0$ (or parameter $\eta$). These
three parameters can in principle be determined by comparing models with 
observations for any remnant, but frequently only one of the postshock
temperatures can be found. Only the postshock electron temperature
$T_{es}$ can be usually determined for young, hot remnants, and only the shock
temperature $T_s$ can be found for old, much colder SNRs. In order to
find both $T_s$ and $T_{es}$ for such SNRs, additional information about
the shock is necessary, such as available from kinematic studies or from
UV and optical observations. It should be possible to determine both $T_s$ and 
$T_{es}$ from the spatially-integrated spectrum alone for SNRs with 
intermediate temperatures and ionization timescales. 

We succesfully explained differences between X-ray spectra for Sedov models
with different parameters in terms of three distribution functions:
the electron temperature distribution, the ionization timescale distribution,
and the ionization-timescale averaged electron distribution. We then considered
a number of simple models, such as a single ionization timescale model and
plane-parallel shock models, and compared their distribution functions 
and X-ray spectra with Sedov models. A frequently used single ionization
timescale model was found to be an inappropriate approximation to Sedov models.
In most situations, plane-parallel shocks provide a much better approximation
to Sedov models, and both plane shocks and Sedov models should be routinely 
used in analysis of SNR X-ray spectra.

The NEI models described in this work, while constituting a basic set of
simplest thermal models, are not sufficient for interpretation of SNR spectra. 
A single plane parallel or a spherical shock is a crude approximation to the
complex morphology of SNRs seen in X-ray images. For complex morphologies, 
observed X-ray spectra are most likely a superposition of shocks with various
velocities, resulting in distribution functions which are poorly described by
a single shock. In general, we do not expect that the NEI models presented 
here can provide perfect or even statistically acceptable fits to SNR spectra,
because of deviations from an idealized plane shock or a Sedov model. This is 
indeed the case for DA 530 (Landecker et al. 1999), 3C397 
(Safi-Harb et al. 2000) and RCW 86 (Borkowski et al. 2000), SNRs whose X-ray 
spectra
were modeled with plane shocks and Sedov models. More effort is clearly
necessary in order to take full advantage of X-ray observations of SNRs. 
Improved modeling must
involve a superposition of shocks with various speeds or determination of
distribution functions directly from observed X-ray spectra, a challenging 
task because of the modest spectral resolution of CCD detectors
onboard {\it Chandra} or XMM-{\it Newton}. 

We based our X-ray emission calculations on Sedov-Taylor dynamics, in
which shocks are not modified by cosmic rays. But if collisionless shocks
were strongly modified in SNRs (Jones \& Ellison 1991), then their dynamics 
would be poorly described
by the Sedov-Taylor self-similar solution, and the resulting X-ray emission
could be very different than in standard Sedov models. Dorfi \& B\"ohringer
(1993) and more recently Decourchelle, Ellison, \& Ballet (2000) showed
that the cosmic-ray modification of shock structure can indeed have 
a dramatic effect on X-ray
spectra of SNRs. This is not suprising because a significant amount of 
energy is deposited into relativistic particles in cosmic-ray 
dominated shocks, which results in lower temperatures of thermal gas. In
several models calculated by Berezhko et al. (1994, 1996), the energy 
fraction contained in cosmic rays may be as high as 80\%\ of the total SNR
kinetic energy.
We found that in such extreme cosmic-ray dominated SNRs the thermal gas
temperature is too low for production of substantial amounts of thermal X-ray
emission, except early in the remnant's evolution when its dynamics and
X-ray emission are strongly affected by SN ejecta. Futhermore, we generally 
found flat temperature profiles in such
models, because of a higher fraction of energy deposited into cosmic rays
at higher shock velocities. Such approximately isothermal SNRs differ 
greatly from Sedov models, which generally contain the hottest gas in their 
interiors. In terms of our distribution functions, the electron temperature
distribution is flat in these isothermal SNRs, the ionization timescale 
is larger because of the increased postshock compression ratio, but the
shape of the ionization timescale distribution is similar to that seen in 
Figure~4. Because of
adiabatic expansion, the ionization-timescale averaged temperature is 
qualitatively the same as depicted in Figures~6 and~7. 

These trends should
be also present in other cosmic-ray modified SNR models with lower
particle acceleration efficiencies. A lower efficiency is actually possible
because the idealized, spherically-symmetric models calculated by Berezhko et 
al. assume a quasiparallel shock everywhere along its periphery, in which 
particle acceleration is most efficient. But realistic SNR blast waves must 
consist of a mixture of quasiparallel and quasiperpendicular shocks, and in 
the absence of strong turbulence, quasi-perpendicular shocks may
not accelerate particles as efficiently. It is then
likely that ``only'' $\sim 50$\%\ of the SNR kinetic energy is transferred
to cosmic rays (Evgeny Berezhko, private communication), in which case 
thermal gas will produce appreciable amounts of X-ray emission during
the late adiabatic evolutionary stage of interest here. But such SNRs must
have significantly different distribution functions than standard Sedov models,
so that they must produce significantly different X-ray spectra. 
A considerable caution is then required when modeling and 
interpreting X-ray observations of SNRs. The distribution functions presented
by us are valid only for Sedov models, without any shock modification by
cosmic rays. It would be worthwhile to carry out an extensive 
investigation of the distribution functions expected in various cosmic-ray 
dominated SNR models, but this task clearly demands a separate effort. 
Note that, unlike for Sedov models, X-ray spectra produced by our 
plane-parallel shocks may be used even for cosmic-ray modified shocks. The
only difference is in interpretation of the shock temperature $T_s$, whose
relationship with the shock velocity $v_s$ is now provided not by the
standard Rankine-Hugoniot shock jump conditions, but by jump conditions
appropriate for a collisionless shock with a particular cosmic-ray 
acceleration efficiency.

Another issue connected with the cosmic ray acceleration in SNRs is the
question of the electron velocity distribution.
In our X-ray emission modeling we assumed that thermal electrons can be
well described by a Maxwellian velocity distribution. This is most
likely a good assumption for the bulk of electrons, usually referred to
as thermal electrons, but the presence of radio synchrotron emission in
SNRs demonstrates that there is a high energy tail to this distribution.
The most energetic ($\sim$ 10 -- 100 TeV) electrons produce 
nonthermal synchrotron 
X-ray emission, as has been recently inferred for SN 1006
\citep{Koyama95}, Cas A \citep{Allen97}, G347.5-0.2
\citep{Slane99}, and RCW 86 (Borkowski et al. 2000), and modeled in detail by
Reynolds (1998). In addition, less energetic electrons in the high energy tail,
with energies of several tens of keV, can also produce lines and continua, 
through the same atomic processes as thermal electrons. (The continuum
produced by such electrons is usually referred as nonthermal bremsstrahlung.)
If these nonthermal processes are important in an SNR, then interpretation of
its X-ray spectrum must rely on a combination of thermal and nonthermal 
models. A good example is SNR RCW 86 where a complex mixture of nonthermal
synchrotron and multitemperature thermal emission is present (Borkowski et 
al. 2000). It is not clear at this time how important nonthermal 
bremsstrahlung is in SNRs, because of the predominance of nonthermal
synchrotron X-ray emission. But if nonthermal bremsstrahlung were important,
energetic electrons producing it would also excite line emission, possibly
leading to changes in line strength ratios and intensities.
The possible presence of such electrons in SNRs is clearly an important issue 
for interpretation of SNR X-ray spectra. One might hope that with the high
quality X-ray observations of SNRs becoming available, we will be able to
determine the distribution functions discussed in this work, and 
unambiguously separate thermal and nonthermal emission in SNRs.

\acknowledgements
Reliable modeling of SNR spectra would not be possible without Fe L-shell
atomic data, which were kindly provided to us by Duane Liedahl. We would
like to thank Andy Szymkowiak, Keith Arnaud, Tim Kallman, and many other
people at Laboratory for High Energy Astrophysics at NASA/Goddard Space
Flight Center who have provided continuous support for this project.
We thank Evgeny Berezhko for discussions about cosmic-ray dominated SNR
models and for making results of his calculations available
to us. We thank Andrej Bykov for stimulating discussions about 
collisionless electron heating. We acknowledge numerous discussions with 
Don Ellison about cosmic-ray modified shocks. Support for this work was 
provided by NASA under grant NAG-7153, and by NSF through a Research
Experiences for Undergraduates award administered by the 
American Astronomical Society.

%\clearpage

%\clearpage

%\clearpage

%\clearpage

%\clearpage

\end{document}